\documentclass[10pt,conference]{IEEEtran}
\IEEEoverridecommandlockouts
\usepackage[numbers]{natbib}
\usepackage{url}

\usepackage{breakurl}
\usepackage[breaklinks]{hyperref}
\usepackage{multirow}
\usepackage{longtable}
\usepackage{amsmath,amssymb,amsfonts}
\usepackage{algorithmic}
\usepackage{graphicx}
\usepackage{textcomp}
\usepackage{xcolor}
\usepackage[varbb]{newpxmath}
\def\BibTeX{{\rm B\kern-.05em{\sc i\kern-.025em b}\kern-.08em
    T\kern-.1667em\lower.7ex\hbox{E}\kern-.125emX}}

\begin{document}

\title{Moving on from the software engineers' gambit: an approach to support the defense of software effort estimates}


\author{
\IEEEauthorblockN{
Patricia G.F. Matsubara\IEEEauthorrefmark{1}\textsuperscript{,}\IEEEauthorrefmark{2}, 
Igor Steinmacher\IEEEauthorrefmark{3}, 
Bruno Gadelha\IEEEauthorrefmark{1}, 
Tayana Conte\IEEEauthorrefmark{1}
\IEEEauthorblockA{\\
\IEEEauthorrefmark{1}Universidade Federal do Amazonas (UFAM), Manaus, AM\\}
\IEEEauthorblockA{\IEEEauthorrefmark{2}Universidade Federal do Mato Grosso do Sul (UFMS), Campo Grande, MS\\}
\IEEEauthorblockA{\IEEEauthorrefmark{3}Northern Arizona University, Flagstaff, AZ, USA\\}
Email: patricia.gomes@ufms.br, igor.steinmacher@nau.edu, \{bruno, tayana\}@icomp.ufam.edu.br}}

\maketitle

\begin{abstract}
Pressure for higher productivity and faster delivery is increasingly pervading software organizations. This can lead software engineers to act like chess players playing a gambit---making sacrifices of their technically sound estimates, thus submitting their teams to time pressure. In turn, time pressure can have varied detrimental effects, such as poor product quality and emotional distress, decreasing productivity, which leads to more time pressure and delays: a hard-to-stop vicious cycle. This reveals a need for moving on from the more passive strategy of yielding to pressure to a more active one of defending software estimates. Therefore, we propose an approach to support software estimators in acquiring knowledge on how to carry out such defense, by introducing negotiation principles encapsulated in a set of defense lenses, presented through a digital simulation. We evaluated the proposed approach through a controlled experiment with software practitioners from different companies. We collected data on participants' attitudes, subjective norms, perceived behavioral control, and intentions to perform the defense of their estimates in light of the Theory of Planned Behavior. We employed a frequentist and a bayesian approach to data analysis. Results show improved scores among experimental group participants after engaging with the digital simulation and learning about the lenses. They were also more inclined to choose a defense action when facing pressure scenarios than a control group exposed to questions to reflect on the reasons and outcomes of pressure over estimates. Qualitative evidence reveals that practitioners perceived the set of lenses as useful in their current work environments. Collectively, these results show the effectiveness of the proposed approach and its perceived relevance for the industry, despite the low amount of time required to engage with it.
\end{abstract}


\section{Introduction}
\label{sec:intro}

A chess opening defines sequences of moves and an overall plan that can impact the game until its end. A now-famous opening is the Queen's Gambit. In a gambit, one player sacrifices material (like a pawn) to gain compensation, such as a gain of tempo or structural weaknesses for the opponent \cite{noauthor_gambit_nodate}. A game-theoretical perspective of software development evinces that software practitioners also play gambits as part of their daily practices. For instance, they can sacrifice the quality of the product to deliver software faster---a sort of software engineers' gambit that leads to technical debt. Along these lines, \citet{vidoni_infinite_2022} proposed a new perspective on technical debt management: an infinite game whose purpose is to continue playing indefinitely instead of winning, i.e., to make the software operational and used for as long as possible.

Considering software development from a game theory perspective, we can explore different strategies to improve our practice \cite{gavidia-calderon_game-theoretic_2020}. In this sense, software development is a game that unfolds in social settings, involving many different players \cite{grechanik_analyzing_2004}: organizations gathering people with varying roles working together, and operating in a market with clients, users, and potential competitors. In such a social context, software effort estimation poses more than just a technical challenge. For instance, pressure is a factor affecting software estimates \cite{matsubara_sextamt_2022}---something unexpected if we consider estimating as a technical prediction task only. Yet, it can lead to the complete rejection of conservative (and accurate) estimates \cite{jones_social_2006} and to arbitrary changes to estimates \cite{magazinius_investigating_2012}, specially when stakeholders suppose developers are not being as productive as they can be \cite{lavallee_why_2015}.

Moreover, poor estimation due to business motivations for earlier deadlines can lead to time pressure in software development \cite{kuutila_time_2020}. Although time pressure can increase motivation and efficiency \cite{kuutila_what_2021}, it negatively affects software practitioners' quality of life, leading to emotions such as sadness and stress \cite{girardi_emotions_2021}. Previous research indicates that the consequences of negative emotions like these include decreasing developers' productivity and increasing delays in executing activities \cite{graziotin_what_2018}, which can put even more pressure on schedules: a hard-to-stop vicious cycle. Time pressure also negatively affects the product quality, making individuals take shortcuts during development, leading to minimal quality assurance tasks, and acting as an obstacle to reviews, among many other quality-related effects \cite{kuutila_time_2020}. In the end, any productivity gains due to pressure probably are insufficient to compensate for the quality costs. It seems the software engineers' gambit is not working. 

Returning to our game theory perspective, all this research shows the results of the strategy of yielding to pressure over software estimates. However, prominent software engineers have long ago proposed an underexplored strategy: the defense of estimates. Notably, \citet{jones_social_2006} emphasized that a relevant reason for unrealistic schedules and time pressure is the inability of software practitioners to defend their estimates. \citet{mcconnell_software_2006} also stressed the need for such a strategy, suggesting the idea of using negotiation principles to implement it.

In this paper, we further contribute to developing the strategy of defending software estimates \cite{jones_social_2006,mcconnell_software_2006,matsubara_best_2022}, supporting software estimators in moving on from the software engineers' gambit. Adopting such a strategy in real-world settings requires a behavior change on the side of practitioners, leading us to enlist behavior change interventions to our aid. Boosts are a promising class of behavioral interventions: they improve competencies, enabling individuals to exercise their agency and empowering them to make better decisions \cite{hertwig_nudging_2017}. Therefore, we propose and evaluate a boost intervention in the form of a digital simulation presenting defense lenses to support the understanding of negotiation principles adapted to the estimation context. The digital simulation takes the format of lightweight interactive videos, with a bit more than 25 minutes of total recorded time, intertwined with a few paused moments when participants choose one action to take in typical pressure scenarios.

Moreover, we are interested in understanding how the digital simulation and the defense lenses can impact software practitioners' behaviors in their daily practices---a concrete step toward Behavioral Software Engineering \cite{lenberg_behavioral_2015}. Thus, we examine whether participation in the digital simulation affects professionals' intentions to defend their software estimates, considering that intentions are the immediate antecedent of behavior, as posited by the Theory of Planned Behavior (TPB) \cite{ajzen_theory_2020}. We also collected data on attitudes, subjective norms, and perceived behavioral control, as these are antecedents of intentions. By analyzing data on TPB, we take advantage of an existing social science theory as part of the foundations of our work, something still surprisingly uncommon in Software Engineering research \cite{lorey_social_2022}.

We carried out an experiment with 32 software practitioners from diverse companies, and with varied experience in software development and maintenance. We randomly assigned the participants to an experimental group participating in the digital simulation or a control one, participating in a brief reflection about pressure over software estimates. 

We found evidence that our approach increased the intentions of software practitioners to adopt the strategy to defend their software estimates. It also increased participants' scores regarding all of the intentions' antecedents. This indicates that participants improved their perceptions that the estimates defense would lead them to outcomes they valued favorably. Moreover, it indicates that participants had more positive beliefs in defense of estimates as a desirable behavior from the perspective of other essential people for them---such as their family, bosses, colleagues, and clients. It also reveals that participants felt more capable of executing it. Participants exposed to our approach were also more inclined to choose a defense action when facing pressure scenarios than the control group. Qualitative data analysis provided evidence that practitioners perceived the set of lenses as useful in their current work environments. Collectively, these results show the effectiveness of the proposed approach and its perceived relevance from the perspective of industry practitioners, even though it takes a low amount of time to engage with it.

\section{Background}
\label{sec:background}

How can we use negotiation to promote the behavior of defending software estimates among software practitioners? To answer this question, we define negotiation and present the methods that formed the foundation for our approach in Section \ref{subsec:negmethods}. Next, we need to understand more about how to model human behavior. We explore the Theory of Planned Behavior (TPB), which we used to model the behavior of defending software estimates, in Section \ref{subsec:TPB}.

\subsection{Negotiation Methods} 
\label{subsec:negmethods}

Negotiation is a back-and-forth communication between two or more parties seeking agreement in something \cite{fisher_getting_2011}. All sides can have shared, opposed, or simply different interests---and they need each other to get desired results \cite{schneider_definition_2017}. A method that profoundly impacted the negotiation teaching \cite{menkel-meadow_why_2006} and practice \cite{tsay_decision-making_2009} is principled negotiation. This method focuses on reaching wise agreements: satisfying legitimate interests of all sides to the best extent, resolving conflicts fairly, and preserving relationships among involved parties \cite{fisher_getting_2011}. In a nutshell, it recommends people to understand (and not succumb to) human nature; to focus on interests instead of people's stated positions; to derive multiple alternative agreement' options; and to use objective criteria to choose among them instead of relying on subjective criteria or pressure.

A complementary method, suited for situations where people are uncooperative and unwilling to reach an agreement, is the breakthrough strategy \cite{ury_getting_2007}. It recommends people to suspend natural reactions when facing a challenging party; to hear instead of arguing during disagreements; to reframe positions into interests; and to use power to educate others based on the consequences of no agreement.

When we are the ones willing to refuse an agreement because it is disadvantageous, we run the risk to make harmful concessions. To avoid this, an alternative method is the positive no \cite{ury_power_2012}. It involves identifying, expressing, and being faithful to our interests. Moreover, it recommends us to look for an alternative course of action we can take independently from others, in case the other side does not accept our no. It also advises us to propose alternatives to the demand or request first made by the other party to keep the relationship and reconcile interests. If needed, we deploy our Plan B respectfully while keeping the door open for a future agreement.

In summary, these methods advise that we do not succumb to emotions when the situation gets complicated, seek to understand the legitimate interest behind positions, look for alternatives to maximize the satisfaction of interests, and focus not only on our side. Other people's situation is also relevant when we wish to preserve relationships. Such knowledge on how to negotiate cooperatively can be instrumental in the estimation context, where people making and receiving estimates are likely to keep their relationships for long periods. Principled negotiation has been proposed in this context earlier \cite{mcconnell_how_1996,mcconnell_politics_2006}, through the creation of tips to estimators. A more comprehensive approach presented based on the exposed methods was also evaluated preliminary \cite{matsubara_best_2022}. In the current work, we build upon these works by creating an approach to support the gain of knowledge about negotiation principles adapted to the estimation context, comprised of a set of defense lenses delivered through a digital simulation. We also assess it through a controlled experiment, gathering evidence on whether it relates to the behavior of defending software estimates. To do so, we needed a theory of human behavior to assist us in understanding such behavior when we expose people to negotiation principles. 

\subsection{The Theory of Planned Behavior}
\label{subsec:TPB}

Among the most used social science theories in Software Engineering research according to \citet{lorey_social_2022}, the Theory of Reasoned Action is the one that focuses on people's overt behaviors \cite{kan_theory_2017}. It has been revised and expanded, leading to the Theory of Planned Behavior (TPB) \cite{ajzen_theory_2020}. Therefore, we chose the TPB to understand more about the behavior of defending software estimates in our study. The TPB affirms the immediate antecedent of behavior is intention, which is a combination of attitudes, subjective norms, and perceived behavioral control regarding the behavior \cite{ajzen_theory_2012}. \citet*{kan_theory_2017} define each of the theory's components:

\begin{itemize}
    \item Behavior is one or more overt actions performed by a person, conceptualized in terms of actions, target, context, and time. It is the object of interest of the researcher. 
    \item Intention is the person's perceived likelihood of performing the behavior. It is the immediate antecedent of the behavior.
    \item Attitude towards the behavior is a person's evaluation (favorably or not) of performing the behavior. It is an aggregate of behavioral beliefs: beliefs that the behavior leads to certain outcomes and how good or bad the individual evaluates such outcomes.
    \item Subjective norm regarding the behavior is a person's perception about how other important people consider the behavior. It is a function of normative beliefs the individual holds regarding what a referent thinks the individual should or should not do. It also includes the motivation the individual has to comply with they believe each specific relevant referent thinks.
    \item Perceived behavioral control is the person's perception of how easy or difficult it is to perform the behavior. It is a function of control beliefs, related to the presence of factors facilitating performance (like resources and abilities) and the absence of factors hindering it (like obstacles).
\end{itemize}

There is no standard questionnaire for TPB studies \cite{ajzen_theory_2020} because each study can focus on entirely different behaviors. Researchers have applied it to varied activities such as time in screen versus time in physical activities \cite{aulbach_dual_2021}, use of transportation alternatives \cite{gardner_going_2010}, and alcohol consumption \cite{cooke_how_2016}, to name a few examples. Nevertheless, there are manuals with instructions for creating TPB-based questionnaires, such as \cite{francis_constructing_2004}. We used the cited manual to construct a questionnaire to assess all TPB variables focused on the behavior of estimators defending (action) estimates (target) of software tasks or projects (context) when facing unreasonable pressure to change them or to accept unrealistic commitments (time).

\section{Approach}
\label{sec:proposal}

Section \ref{subsec:design} describes our approach for the design of our artifacts. Next, Section \ref{subsec:lenses} describes the defense lenses and Section \ref{subsec:ds} describes the digital simulation. 

\subsection{The Approach Design}
\label{subsec:design}

We designed our proposed solution using a Design Science Research (DSR) approach \cite{hevner_three_2007}. In summary, we conducted two studies to investigate our research problem. The first was a systematic literature mapping of factors affecting software estimates when using expert-judgment \cite{matsubara_sextamt_2022}. The second was a qualitative study in industry, investigating how practitioners transform estimates into commitments \cite{matsubara_buying_2021}. These two studies evidenced the problem of pressure over estimates, leading practitioners to either change their estimates to more ``acceptable'' ones to stakeholders or to use padding of other tasks/projects to compensate for unrealistic commitments. Therefore, they are part of our DSR relevance cycle---although their results also represent contributions related to the rigor cycle.

Next, as part of our DSR rigor cycle, we investigated the negotiation literature to form our theoretical foundations. After extensively studying the negotiation methods Section \ref{subsec:negmethods} describes, we designed the first version of the defense lenses as part of our design cycle. We ran a focus group study with software practitioners to gather preliminary evidence on the lenses' perceived usefulness, and improvement opportunities \cite{matsubara_best_2022}, also as part of our design cycle. Practitioners did not understand some of the lenses, required more examples of their application, and asked for an alternative presentation format. We realized we could address these issues by creating another artifact to support the lenses, so we designed the digital simulation and assessed it through the controlled experiment we present here. The controlled experiment is part of the DSR rigor cycle.

\subsection{The Defense Lenses}
\label{subsec:lenses}
 
The negotiation principles were adapted to the estimation context and embodied in the format of defense lenses. Table \ref{tab:defenseLenses} present the whole set of lenses, the pack that the lens belongs to, and a brief description of each lens.

\begin{table*}[htpb]
\caption{Complete set of defense lenses.}
\label{tab:defenseLenses}
\begin{tabular}{l|p{3cm}|p{11cm}}
\textbf{Pack} & \textbf{Lens} & \textbf{Brief description} \\ \hline
Minimal & Assert your estimate & Show that the estimate satisfies the interests and needs of everyone involved and how changing it does more harm than good.\\ \hline
Minimal & Laddering whys & Investigate the most profound reasons that lie under the estimate, getting to the basic human needs that justify it---and that one wishes to protect. The deeper one digs into one's interests, needs, and values, the more likely one will hit a bedrock on which to stand firm. \\ \hline
Minimal & Pressuring forces & Investigate what underlies the pressure over the estimate to understand the perspective of the person making pressure. Go beyond their position to get their deepest interests. \\ \hline
Minimal & Candidate commitments & Go beyond deflecting pressure over estimates to propose a third option: a commitment that aims at satisfying both sides to the greatest extent possible. \\ \hline
Minimal & Choose your battles & Be honest to oneself to identify situations when keeping the estimate does not truly advance one's deepest interests and needs. It brings a bit of strategic thinking to the tense moment of enduring pressure. \\ \hline
Extended & Keep strategy & Clarify that one would change the estimate if possible, but there are legitimate reasons for keeping it. Sometimes, such reasons are outside one's control. \\ \hline
Extended & Perspective taking & Recognize when people adopt pressuring tactics to neutralize their effects effectively. \\ \hline
Extended & Reality test & Ask reality-testing questions to educate people about the consequences of changing the estimate. Asking people what will happen is better than telling it. \\ \hline
Extended & Golden bridge & Sometimes, people reject the estimate because they have no alternative. Show them the connection between their needs and the estimate to reach an agreement. \\
\end{tabular}
\end{table*}

The Minimal Pack contains the core lenses for estimators to handle concrete pressure episodes over their estimates in the short run. The pressure episode can happen during a group estimation session or when an estimator provides an individual estimate directly to a client or a higher manager. The other pack is titled the Extended Pack and complements the first one. It is helpful when estimators have tried to defend their estimates, but the pressure continues. All the lenses are in a booklet, which is part of the supplementary material \cite{matsubara_moving_2023}. 

\subsection{The Digital Simulation}
\label{subsec:ds}

\begin{figure*}[htb]
  \centering
  \includegraphics[width=\textwidth]{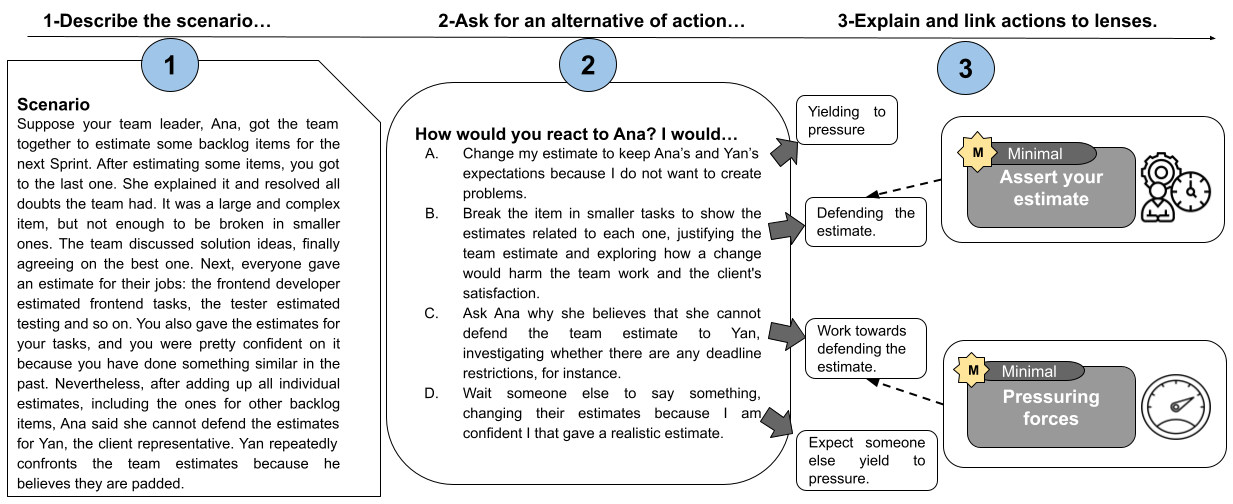}
  \caption{The interactive video dynamics.}
  \label{fig:video}
\end{figure*}

To aid software practitioners in defending their software estimates, we designed a digital simulation to disseminate negotiation principles adapted to the estimation context. Digital simulations are technology-based simulations that model a process or a system \cite{gegenfurtner_digital_2014}. They provide opportunities to adjust aspects of reality to facilitate learning and practice in varied ways, such as by addressing infrequent events or by enabling immediate feedback on the learners' actions \cite{chernikova_simulation-based_2020}. We implemented the digital simulation as interactive videos with pressure scenarios and embedded questions, using the platform PlayPosit\footnote{\url{https://playpos.it/}}. We prepared two videos: one for each pack. Figure \ref{fig:video} illustrates the video dynamics.

After a brief introduction, the video described a realistic written pressure scenario. Figure \ref{fig:video}.1 shows the first scenario presented in the video about the Minimal Pack. Following, we asked the participant which action they would choose to respond to the scenario of a set of four alternatives. The video paused to allow participants to think. Alternatives included one or two options that represented yielding to pressure---while all others represented paths towards defending the estimates. In Figure \ref{fig:video}.2, we can see all the options for the specific scenario illustrated in the image. Options A and D represent yielding to pressure or expecting someone else to do it, respectively. After the participant chose the options they wanted, the tool presented a score to indicate whether the answers aligned with the lenses' ideas. For instance, if participants choose options B and C, they get a 100\% score. If they choose only one of them, they get a score of 50\%. Otherwise, they get a zero score. Next, we discussed each alternative, identifying the ones that represented concessions and connecting the other ones to the defense lenses that supported the depicted behavior. Therefore, as Figure \ref{fig:video}.3 presents, we identified to participants that options A and D represent yielding to pressure. Then, we explained the lens ``Assert your estimate'' in light of option B and the lens ``Pressuring forces'' in light of option C.

\section{Experimental Design}
\label{sec:method}

In this study, our goal is to support software estimators in defending their software estimates when facing pressure to change them or to accept unrealistic commitments. To guide our research efforts towards its satisfaction, we addressed the following research questions: (i) RQ 1 - Does the participation in the digital simulation increase software practitioners' intentions to defend their software estimates, as well as its antecedents (attitudes, subjective norms, and perceived behavioral control)? and (ii) RQ 2 - What is the perceived usefulness of the defense lenses in real-world situations from the perspective of participants of the digital simulation?

To answer them, we carried out a controlled experiment including an experimental and a control group. We exposed the experimental group to the digital simulation and to the defense lenses as described in Section \ref{sec:proposal}, expecting it to stimulate changes in the intentions of defending software estimates. For comparison, we exposed a control group to reflection questions about past pressure scenarios they faced in their jobs and the impact of such pressure. If we can get higher intentions to defend software estimates from participants in the control group, we would have a much simpler intervention to propose for estimators. Thinking of pressure scenarios and their consequences immediately before communicating estimates would be easier and cheaper than studying the defense lenses through the digital simulation.

We collected data in pre- and post-questionnaires based on the Theory of Planned Behavior and on the reaction of participants to a set of pressure scenarios, as described in more detail in Section \ref{subsec:collection}. We piloted the questionnaires with four participants: two in the control and two in the experimental group. We improved the wording of a few items. During the pilot study, we also included the reflection questions in the post-questionnaire of the experimental group---a decision we changed for the final study. Moreover, we did not include the pilot participants' data in our final analysis.

Therefore, considering the previous literature suggesting that we can use negotiation principles in defense of software estimates, we hypothesized that after participating in the digital simulation and learning about the defense lenses, participants in the experimental group would exhibit:

\begin{itemize}
    \item ...\textbf{higher levels of attitude} to defend their software estimates than before participating (H1a) and than participants in the control group (H1b). 
    \item ...\textbf{higher levels of subjective norm} to defend their software estimates than before participating (H2a) and than participants in the control group (H2b).
    \item ...\textbf{higher levels of perceived behavioral control} to defend their software estimates than before participating (H3a) and than participants in the control group (H3b).
    \item ...\textbf{higher levels of intention} to defend their software estimates than before participating (H4a) and than participants in the control group (H4b).
\end{itemize}

In addition, we assessed the participants' reaction to pressure scenarios we designed inspired by the SE literature, believing participants in the experimental group were more likely to pick alternatives representing the defense of estimates. Thus, we also hypothesized that:

\begin{itemize}
    \item After participating in the digital simulation and learning about the defense lenses, participants in the experimental group will choose more defense actions in pressure scenarios than participants in the control group (H5a).
\end{itemize}

For each of these research hypotheses, we have a corresponding null hypothesis, stating that after participating in the digital simulation, participants in the experimental group would exhibit \textbf{lower or equal levels} of each given variable than before participating (for H1a-H4a) and than participants in the control group (for H1b-H4b). For instance, the corresponding null hypothesis for H1a states that participants in the experimental group exhibit \textbf{lower or equal levels} of attitudes to defend their software estimates than before participating. Also, the corresponding null hypothesis for H5a states that after participating in the digital simulation and learning about the defense lenses, participants in the experimental group will choose an \textbf{equal number or fewer} defense actions in pressure scenarios than participants in the control group.

\subsection{Sampling Strategy and Participants}
\label{subsec:participants}

We invited 75 people with varying experience in software development and maintenance, both in terms of roles and years of work, from our network to participate in the experiment. All of them were involved with the estimation of software tasks or projects. A total of 45 people accepted to participate in the study. We randomly assigned them to the experimental and control groups after they answered the pre-questionnaire. We had 23 participants assigned to the control group and 22 to the experimental group. From these, 32 participated in all study stages: 16 in the control group and 16 in the experimental group. Participants that dropped out of the study were more experienced and more advanced in their careers. They indicated they had trouble finding time to participate. Table \ref{tab:demographics} presents the descriptive statistics for the demographics of participants in each group.

\begin{table}[htp]
\caption{Demographics of participants.}
\label{tab:demographics}
\begin{tabular}{p{1.1cm}|p{3.2cm}|p{3.2cm}}
 & \textbf{Control} & \textbf{Experimental} \\
  & n = 16 & n = 16 \\ \hline
Gender & Men = 11  & Men = 13 \\ 
 & Women = 4  & Women = 3 \\ 
 & Other = 1 & Other = 0 \\ \hline
Age & mean = 30.6 (sd = 5.6) & mean = 30.1 (sd = 5.0) \\ \hline
Experience & mean = 8.1 (sd = 4.9) & mean = 7.3 (sd = 4.8) \\ \hline
Education level & High School = 1 & High School = 1 \\
 & Bachelor's/College Degree = 12 & Bachelor's/College Degree = 9 \\ 
 & Masters or PhD Degree = 3 & Masters or PhD Degree = 6 \\ 
\end{tabular}
\end{table}

The control group participants had almost one year more experience in software development and maintenance than the participants in the experimental group on average. Visual inspection of the data reveals the control group is slightly more experienced---but the difference was not statistically significant. As for educational level, the experimental group had a few more participants with at least a Master's Degree than the control group. We also collected data from participants' roles. Around 90\% of the participants in the experimental group and 80\% in the control group reported that they work as Developers, Machine Learning Engineers/Data Scientists, Tech Leads, or Agile Experts. The rest of the participants reported being managers, product owners, or requirements engineers. 

Collectively, the study participants belonged to a total of 32 different organizations. The companies were mostly medium to very large and had five years or more of existence, i.e. were not so young. Regarding industry, most were from IT services and consultancy businesses. However, there were companies from banking, health services, financial services, oil and gas, retail, insurance, government, media production, and distribution, among others.

\subsection{Data Collection}
\label{subsec:collection}

To collect data on the intention to defend software estimates and the other TPB variables, we built a questionnaire following the instructions in \citet{francis_constructing_2004}. We derived 15 questions: three for intentions and four for each of intentions' antecedents (attitudes, subjective norms, and perceived behavioral control). All items follow a seven-point Likert scale with neutral. For instance, one of the items regarding perceived behavioral control was: ``I am confident that I could defend an estimate when facing unreasonable pressure if I wanted to'', with options going from ``strongly disagree'' to ``strongly agree''.

To assess the reaction of software practitioners, we also derived scenarios representing situations with pressure over software estimates. In each scenario, the participant had to choose one of four alternative action options---where one always represents the behavior of defending the software estimates. In contrast, the others represent the behavior of yielding to pressure. The complete list of questions and scenarios is part of our supplementary material \cite{matsubara_moving_2023}. 

Both groups answered the first questionnaire in the first moment (M1), with demographic and TPB questions. We sent the questionnaire via e-mail to all participants, asking them to answer it within one week. In the second moment (M2), the operationalization of the study changed for each group. 

Participants in the experimental group engaged with the digital simulation and were exposed to the defense lenses. We sent them the links to the interactive videos, giving them one week to watch. At the end of the video, we left a link for the final questionnaire, with the same questions regarding TPB from M1 plus questions about the actions they would take in the five pressure scenarios. We also asked whether they considered the lenses and negotiation principles would help deal with pressure in their work. If they answered ``yes'', we also asked which lenses or principles they considered the most useful and why.

In M2, participants in the control group answered questions about pressure scenarios they faced in their jobs and what was their typical outcomes. Next, we asked them to answer the TPB and scenario questions. By doing so, we expected to create a priming effect regarding past pressure experiences from participants. Primes are used in psychology research to selectively increase the accessibility of specific conceptions or pieces of information in memory \cite{bless_assimilation_2016}, leading to changes in behavior. For instance, past research has shown that we can prime power in applicants for jobs, making them feel either powerful or powerless immediately before the writing of application letters or interviews, and either improving or worsening their application outcomes, respectively \cite{lammers_power_2013}. Therefore, we expected the past pressure scenarios and their outcomes to prime the typical behavior of participants in such situations, creating a baseline behavior for comparison.

\subsection{Data Analysis}
\label{subsec:analysis}

In our data analysis, we employed the traditional Null Hypothesis Significance Testing (NHST). However, the Software Engineering research community gets increasingly aware of its limitations---such as the rejection of the null hypothesis based on the probability $\mathbb{P}[data|H\textsubscript{0}]$ (p-values), when we need the posterior probability $\mathbb{P}[H\textsubscript{0}|data]$ to accept or reject a hypothesis based on empirical data \cite{furia_what_2017}. Moreover, with NHST there is a higher chance of rejecting the null hypothesis as the number of observations grows because it is usually more restricted than the alternative hypothesis. So it gets likelier that some effect is detectable \cite{furia_bayesian_2021}. The research community has proposed using Bayesian Hypothesis Testing as an alternative \cite{erdogmus_bayesian_2022}. Thus, we also adopted it to compare the plausibility of the research hypotheses and null ones \textit{relative to one another}.

We tested whether $M2varE > M1varE$, i.e., we assessed whether the variable of interest (attitude, subjective norm, perceived behavioral control, or intentions) was higher for the \textit{experimental group} at Moment 2 (post-questionnaire) compared with Moment 1 (pre-questionnaire). We used the Wilcoxon Signed-Rank paired samples test for that. We also assessed whether $M2varE > M2varC$, i.e., we tested whether the variable of interest at Moment 2 was higher for the \textit{experimental group} than for the \textit{control group}. In this case, we used the Mann-Whitney test. In all cases, we applied one-sided testing because previous literature supports the idea that negotiation principles could aid in defending software estimates---a reason to believe that results would increase in the experimental group after the intervention and when compared with the control group.

Furthermore, we carried out a reliability analysis for the TPB questionnaire. We dropped one item for attitudes, one for subjective norms, and one for perceived control to improve Cronbach's $\alpha$ score. We give more details on this in Section \ref{sec:threats} and the supplementary material \cite{matsubara_moving_2023}.

\section{Results}
\label{sec:results}

So, does participation in the digital simulation and learning about the defense lenses increase software practitioners' intentions to defend their software estimates, as well as its antecedents? We focus on answering this first question in Section \ref{subsec:intentionsResults}. Also, what is the perceived usefulness of the defense lenses in real-world situations from the perspective of participants of the digital simulation? We focus on this second question in Section \ref{subsec:usefulnessResults}.

\subsection{RQ 1: Intentions, its Antecedents, and Pressure Scenarios}
\label{subsec:intentionsResults}

\begin{figure*}[htb]
  \centering
  \includegraphics[width=1\textwidth]{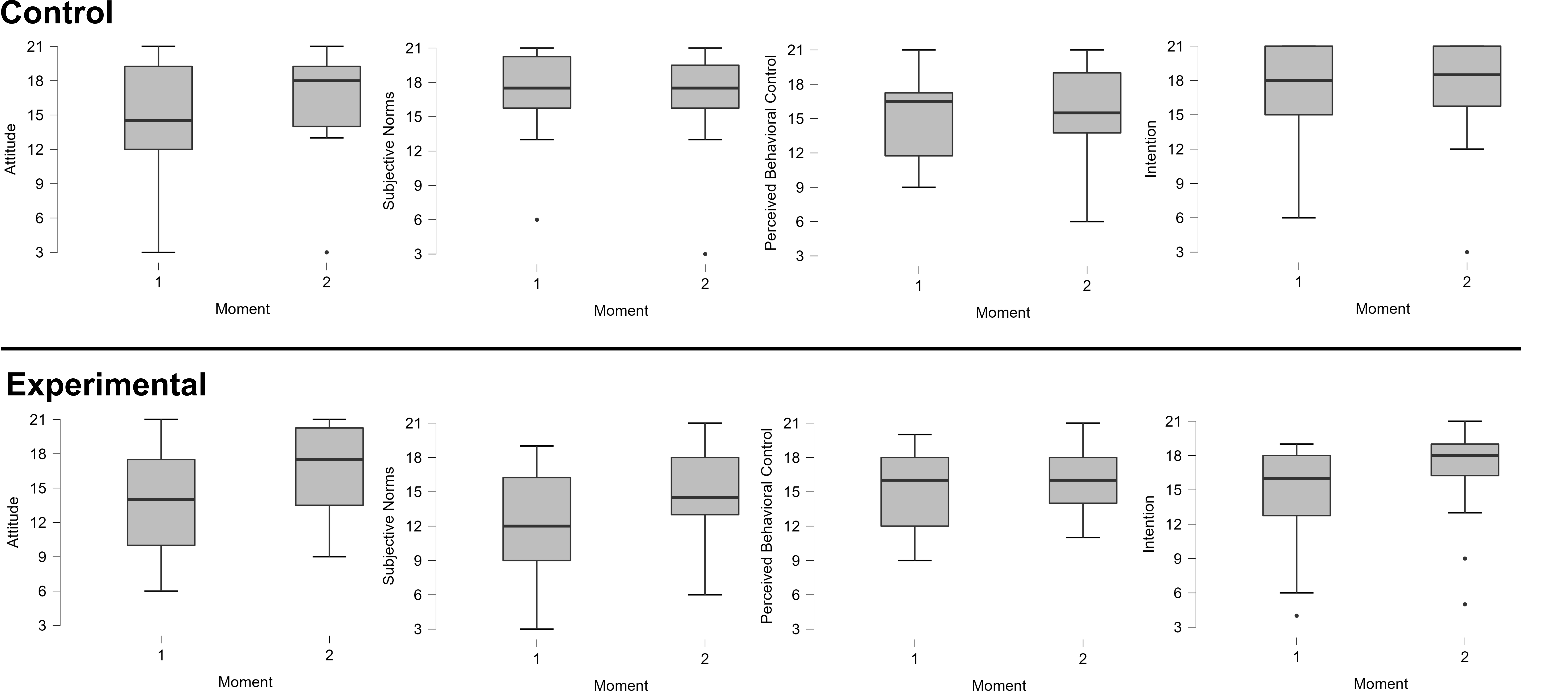}
  \caption{Boxplots for Theory of Planned Behavior' variables. 1 = before the intervention; 2 = after. Generated with JASP (\url{https://jasp-stats.org/}).}
  \label{fig:boxplots}
\end{figure*}

As Section \ref{sec:method} describes, we investigated the impact of the digital simulation exposing the defense lenses compared with reflecting on pressure scenarios over the intentions of estimators to defend their software estimates. In Figure \ref{fig:boxplots}, we present the boxplots of the TPB variables of the experimental and control group before and after the intervention. 

Overall, the values for all variables improved, except for the control group's subjective norms and both groups' perceived behavioral control. Regarding the choice of defense actions in pressure scenarios, the experimental group chose four out of five (median), and the control group chose three out of five. To verify the statistical significance of these results, we employed Null Hypothesis Significance Testing (NHST), which is the frequentist approach to data analysis. We tested normality for all variables using the Shapiro-Wilk test. In the within-group test, perceived behavioral control and intentions were not normal. In the between-groups test, attitudes, subjective norms (for the experimental group), intentions, and the number of chosen defense actions in scenarios were not normal. Therefore, we used the Wilcoxon Signed-Rank test for the paired sample (Hypotheses H1a-H4a) and the Mann-Whitney test for the independent samples (Hypotheses H1b-H5b). As we explained in Section \ref{subsec:analysis}, we also carried out a Bayesian Hypothesis Testing (BHT) to compare the plausibility of the research and null hypotheses \textit{relative to one another}. In Table \ref{tab:NHSTres}, we present the results for NHST and BHT analyses.

\begin{table*}[htpb]
\caption{Analysis Results. RBC = Rank-Biserial Correlation.}
\label{tab:NHSTres}
\begin{tabular}{p{3.6cm}|p{1cm}|p{1cm}|p{1.5cm}|p{2cm}|||p{1.5cm}|p{1.5cm}|p{1.5cm}}
& \multicolumn{4}{c|||}{Null Hypothesis Significance Testing} & \multicolumn{3}{c}{Bayesian Hypothesis Testing} \\ \hline
\textbf{Hypothesis} & \textbf{W} & \textbf{p-value} & \textbf{RBC} & \textbf{Interpretation} (Statistical Significance) & \textbf{BF\textsubscript{+0}} & \textbf{Effect Size} (Median) & \textbf{Interpretation} (Strength of Evidence) \\ \hline
H1a: attitudes in exp. group & 90.0 & 0.047 & 0.500 & Significant & 1.370 & 0.374 & Weak \\ \hline
H2a: subjective norms in exp. group & 73.5 & 0.027 & 0.615 & Significant & 5.317 & 0.555 & Moderate  \\ \hline
H3a: PBC in exp. group & 72.5 & 0.109 & 0.381 & Non significant & 1.394 & 0.378 & Weak \\ \hline
H4a: intentions in exp. group & 91 & 0.008 & 0.733 & Significant & 5.319 & 0.556 & Moderate \\ \hline \hline

H1b: attitudes between groups & 124.5 & 0.560 & -0.027 & Non significant & 0.302 & 0.201 & Moderate \\ \hline
H2b: subjective norms between groups & 82.5 & 0.959 & -0.355 & Non significant & 0.160 & 0.114 & Moderate \\ \hline
H3b: PBC between groups & 126.0 & 0.538 & -0.016 & Non significant & 0.329 & 0.201 & Moderate \\ \hline
H4b: intentions between groups & 100.5 & 0.858 & -0.215 & Non significant & 0.201 & 0.135 & Moderate \\ \hline \hline

H5a: scenarios between groups & 168.5 & 0.052 & 0.316 & Non significant & 1.219 & 0.432  & Weak \\
\end{tabular}
\end{table*}

\begin{figure*}[htb]
  \centering
  \includegraphics[width=1\textwidth]{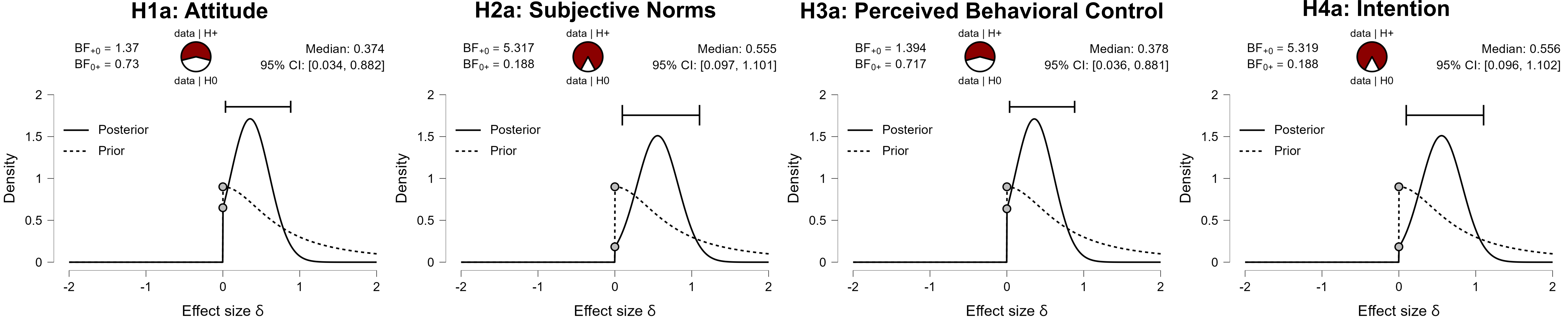}
  \caption{Hypothesis Testing Results for the Experimental Group Before and After the Intervention. Generated with JASP (\url{https://jasp-stats.org/}).}
  \label{fig:HTResultsHa}
\end{figure*}

 Figure \ref{fig:HTResultsHa} presents the results for the experimental group before and after they engaged with the digital simulation with the defense lenses. In the figure, BF\textsubscript{+0} is the Bayes Factor for our research hypotheses (H1a-H4a), which predicted improvements in all variables. At the same time, BF\textsubscript{0+} is the Bayes Factor for the null hypotheses representing no improvement. The Bayes Factor tells us the extent to which a hypothesis predicts the given data compared to others, providing a measure of the strength of evidence of one over the other \cite{erdogmus_bayesian_2022}. The figure also presents a visual representation of the Bayes Factor through a pie chart---the larger the red area, the higher the support for the research hypotheses. The dotted density line represents the prior, which in our case was uninformative (a Cauchy distribution with a scale equal to 0.707), while the full density line presents the posterior. The gray dots represent the prior and posterior specific densities at the test value. When the gray dot of the posterior gets far below the one for the prior, we have higher support for the research hypothesis. The figure also shows the median of the effect size and its 95\% confidence interval (CI)\footnote{This is the effect size on a latent level (see \citet{van_doorn_bayesian_2020})}.

Figure \ref{fig:HTResultsHa} reveals that our research hypothesis regarding attitudes after exposure to our approach is a bit more probable than the null hypothesis (due to the BF\textsubscript{+0} of 1.370). Therefore, although we have evidence in favor of H1a, it is weak due to a Bayes Factor between one and three\footnote{We used the ranges of the strength of evidence of \citet{erdogmus_bayesian_2022}.}. Figure \ref{fig:HTResultsHa} also shows that the intervention impacted subjective norms in the experimental group, with a BF\textsubscript{+0} of 5.317, which is evidence of moderate strength in favor of H2a. We also have evidence of weak strength of an effect over perceived behavioral control in Figure \ref{fig:HTResultsHa}, given the BF\textsubscript{+0} of 1.394. Figure \ref{fig:HTResultsHa} indicates that the intervention also impacted intentions, with a BF\textsubscript{+0} of 5.319, showing we have evidence of moderate strength in favor of H4a over the null hypothesis. Therefore, in light of the Theory of Planned Behavior, these results indicate that within the experimental group, the proposed approach increased the intentions of defending software estimates when facing pressure, probably through improvements in all its antecedents, primarily through subjective norms. 

Regarding the hypotheses comparing the control and experimental groups, we found weak to moderate evidence only for the null hypotheses for all the TPB variables, with BF\textsubscript{0+}\footnote{Reminder: BF\textsubscript{0+} is the Bayes-Factor for the null hypotheses.} varying from 3.01 for attitudes to 6.23 for subjective norms. Therefore, we found no support from evidence for H1b-H4b, as we found no evidence of higher scores for the experimental group. As for the response to pressure scenarios, we found a BF\textsubscript{+0} of 1.219, which favors H5b but with evidence of weak strength. All the graphs for the testing regarding Hypotheses H1b-H5b are part of our supplementary material \cite{matsubara_moving_2023}.

Mostly, the BHT and Null Hypothesis Significance Testing (NHST) analyses agree. One exception is regarding ``attitudes" in the experimental group. Results are statistically significant (p-value = 0.047) for NHST. This would lead us to reject the null hypothesis and consider there is an effect. However, the BHT reveals that the evidence in favor of the research hypothesis is weak. The BF+0 = 1.370 value means that the observed data in our experiment is only 1.370 times more likely to happen under the research hypothesis than the null one \cite{erdogmus_bayesian_2022}. Instead of rejecting the null hypothesis, we conclude that the research hypothesis does not stand out as a strong explanation of our data for this variable, and we are not confident to say there is an effect of the digital simulation on participants' attitudes. This is a good example of how relying on p-values alone might increase the chances of a Type-I error.

In summary, considering both analyses, the research hypothesis as a better explanation of the data than the null hypothesis only for the variables of subjective norms and intentions in the experimental group. Or, in the frequentist interpretation: we reject the null hypothesis only for these two variables.

\subsection{RQ 2: Perceived Usefulness}
\label{subsec:usefulnessResults}

As part of the post-questionnaire, we asked participants of the experimental group about their perceived usefulness of learning negotiation principles encapsulated in the defense lenses in the digital simulation. If they answered positively, we asked them to explain which lenses or principles they believed to be the most useful for defending software estimates in their work and why. Otherwise, we asked them why the lenses would not be helpful.

All participants stated they found the principles can be useful in their work environments. Different participants mentioned different lenses/principles as the most useful ones. For instance, P5 chose the ``Assert your Estimate'' lens ``because it exercises the appreciation of the rationale that lies behind the estimate''. P18 picked the Laddering Whys lens as ``asking questions makes us reflect on motivations and difficulties of people involved in the projects, and I think this will be useful in my estimates.'' P29 chose the ``Candidate Commitments'' lens because ``in my experience, flexibility is really important in negotiations. Trying to find a commitment that brings benefits to all parties is hard, but I believe is the best alternative.'' All these lenses were from the Minimal Pack, but some participants also chose lens from the Extended Pack too. For example, P2 pointed the ``Keep Strategy'' lens as the most useful one ``because I always have reasons outside my control, so I can look for tools to defend my estimate.'' Another participant, P11  chose the ``Perspective Taking, because when we take a wider look we can have arguments to base or to balance a point of view.''

Collectively, participants mentioned all lenses, except one: ``Choose your battles''. The most cited lens was the Candidate Commitments one. Moreover, some participants stated that all lenses are relevant because they complement each other, and each lens fits a different situation. In the words of P35: \textit{``I think the principles were interesting. It is possible to balance time x quality. I believe the lenses complement each other, and knowing how to negotiate is part of the job''}. One participant also stated that both videos helped gain confidence to defend the software estimates. Next, we discuss these results in-depth.

\section{Discussion}
\label{sec:discussion}

Are software engineers willing to move on from the software engineers' gambit in the software development game? In the following sections, we focus on this.

\subsection{Comparing the Groups}

We found weak to moderate evidence in favor of the null hypotheses of no increase in any of the TPB variables (intentions, attitudes, subjective norms, and perceived behavioral control) for participants in the experimental group \textit{compared with} participants in a control (Section \ref{subsec:intentionsResults}). To understand more about this, we tested the difference in scores between these groups \textit{before} the interventions. We found evidence of moderate strength for differences in subjective norms (BF\textsubscript{10} = 5.010) and of weak strength for intentions (BF\textsubscript{10} = 1.390) \cite{matsubara_moving_2023}. This indicates that participants in the control group were possibly more inclined to defend their estimates in the first place. The evidence we have in favor of the proposed approach when comparing the two groups comes from the response to pressure scenarios: participants in the experimental group chose action defenses in one scenario more (median) than the control group. This is a positive answer to the question that opened this section, showing a higher willingness to adopt the strategy of defending software estimates when participants learned more about it.

\begin{table}[h!]
\begin{tabular}{|p{8.4cm}|}
\hline
\textbf{Takeaway message:} Exposing software practitioners to a digital simulation presenting the defense lenses with negotiation principles increases their chances of defending their estimates instead of yielding to unreasonable pressure. \\ \hline
\end{tabular}
\end{table}

\subsection{Before and After the Intervention - Experimental Group}

Another part of the answer to the question opening this section comes from the difference before and after the intervention, considering experimental group participants. We found weak evidence of higher scores for attitudes. Attitudes are a function of beliefs about a behavior's likely consequences---their outcomes or experiences \cite{ajzen_theory_2020}. The evaluation of such outcomes---as desirable or not--- also matters \cite{kan_theory_2017}. In our study, the digital simulation probably improved the participants' beliefs that performing the defense of estimates would lead to outcomes that participants regard favorably. This can include better work experience, sufficient time to make higher-quality deliveries, and lower overtime work.

We also found evidence of higher scores for subjective norms, which refer to the person's perceptions about (i) whether relevant referent individuals or groups approve (or disapprove) the behavior in question and (ii) whether such referents perform it \cite{ajzen_theory_2020}. The referent's importance to the person also plays a role \cite{kan_theory_2017}. So the digital simulation might have caused participants to think that colleagues, bosses, and clients would approve of defending estimates, given that it can protect product quality, the company image, and other of their interests.

It was surprising to find evidence of only weak strength for improvements in perceived behavioral control, as it is about beliefs regarding the presence of factors that can facilitate (or hinder) the behavior performance, including beliefs about skills \cite{ajzen_theory_2020}. This result was not a matter of participants misunderstanding the lenses: when discussing their perceived usefulness, participants demonstrated understanding correctly the principles the lenses embodied. One possible explanation relates to the measurement instrument: it might not have covered all relevant items for perceived behavioral control. We explore this issue in Section \ref{sec:threats}. Another explanation is that participants may need more time to exercise their newly acquired knowledge. Previous research on digital simulations suggested that they can make participants more aware of what they do not know, realizing their increasing skills as the training goes on for a more extended period \cite{gegenfurtner_digital_2014}. Yet another possible explanation is that impeding control factors, such as lack of cooperation from other stakeholders, might play a large role in the control feelings regarding this specific behavior. Maybe the defense lenses learned through the digital simulation might look like a tool that does not directly affect such factors at first glance. However, the defense lenses are primarily about dealing with people unwilling to cooperate to define a realistic commitment based on estimates. Again, understanding this might require more experience using the lenses, possibly through longer learning periods. In the following takeaway message, we use the term ``very likely'' because of the moderate strength of evidence. For evidence with weak evidence, we use the term ``likely''.

\begin{table}[!h]
\begin{tabular}{|p{8.4cm}|}
\hline
\textbf{Takeaway message:} Exposing software practitioners to a digital simulation presenting the defense lenses with negotiation principles is very likely to improve their perceptions about subjective norms towards the behavior of defending their software estimates. \\ \hline
\end{tabular}
\end{table}

Regarding intentions, we found evidence of moderate strength of an increase in scores: a clear indication that people got more inclined to defend their estimates after the digital simulation instead of succumbing to the software engineer's gambit. The stronger the intention, the more likely people will perform the behavior. However, it is no assurance as a varied set of factors can prevent people from acting on their intentions \cite{ajzen_theory_2020}. In any case, we have evidence that a short digital simulation---the two videos are no longer than 30 minutes together---is enough for people to grasp the ideas behind the set of lenses and positively change their intentions. 

Qualitative analysis of the perceived usefulness of the lenses also provides evidence in favor of the applicability of the lenses in the field. Collectively, participants mentioned all the lenses as useful except for one: the ``Choose your battles'' lens. Interestingly, this lens aims to help estimators to act strategically, identifying the situations that keeping the estimate might not be in their best interest---or in other words: when yielding to pressure might be the best option.

\begin{table}[!h]
\begin{tabular}{|p{8.4cm}|}
\hline
\textbf{Takeaway message:} Exposing software practitioners to a digital simulation presenting the defense lenses with negotiation principles is very likely to increase their intentions to perform the behavior of defending their software estimates. \\ \hline
\end{tabular}
\end{table}

\subsection{Before and After the Intervention - Control Group} 

Another side of our answer regards the difference before and after the reflection questions in the control group. We expected the reflection on the pressure scenarios and their impact on participants' lives, projects, and organizations could elicit the typical behavior people have when facing pressure over their estimates. Thus, we tested whether this simpler intervention of reflection, which takes much less to participate in than the digital simulation, would also increase scores for intentions. We executed the same analysis procedures we did for the experimental group. We found weak evidence in favor of improvements only for attitudes (BF\textsubscript{+0} of 2.403). This shows that the reflection also makes evident how good and useful the defense of estimates is, possibly through remembering how bad pressure outcomes are. However, the evidence favored the null hypotheses for all other variables, as we fully report in the supplementary material \cite{matsubara_moving_2023}. Furthermore, descriptive statistics suggest a drop in perceived behavioral control. Therefore, while attitudes increased through reflection, the feeling of the capacity to perform the behavior decreased.

\begin{table}[!h]
\begin{tabular}{|p{8.4cm}|}
\hline
\textbf{Takeaway message:} Exposing software practitioners to a reflection on pressure scenarios and their outcomes is likely to improve attitudes towards the defense of software estimates, without leading to a corresponding increase in intentions of performing such behavior. \\ \hline
\end{tabular}
\end{table}

\subsection{Cost x Benefits of the Digital Simulation} 

The final issue to discuss about moving on from the software engineers' gambit is the practical significance of our results \cite{torkar_method_2022}: the core concern in supporting the choice of concrete actions by practitioners in the software industry. After all, is it worth engaging with the digital simulation and the study of the defense lenses? To answer that, we need to analyze the costs and benefits of doing so. Starting with the \textbf{cost}, it takes less than 30 minutes to participate in the digital simulation. For the ones interested in reading the booklet with the defense lenses, it can take up to 30 minutes more\footnote{Approximately 13 minutes for the main text and one minute and a half for each lens. We made this estimation using \url{https://thereadtime.com/} for silent reading (around 238 words per minute).}.

Assessing the \textbf{benefits} is not so straightforward. First, what are the expected outcomes of an increased intention to defend realistic software estimates, considering this is the main result of participating in the digital simulation? We expect it to increase defense behavior, but it remains unclear how much of an increase in intention is necessary to secure that. In any case, when asked whether they think learning negotiation principles is useful for defending software estimates in their current work environment, all participants of the experimental group answered positively. This can be a rough indication that such people will follow on their intentions to defend their estimates. As revealed in the words of one of the participants (P22), sent to us through an e-mail after participation in the experimental group: \textit{``The lenses are broad concepts and have good application in the real world. And I think they are applicable in any relation of estimates x delivery. (...) I am interested in going deeper on this and delivering this content to my manager. He is the one who gets our estimates and presents them to the client. (...) It would also be nice to apply the lenses when developers are estimating.''}

Second, if an increase in defense behavior does materialize, what are the consequences for individuals and companies? The qualitative analysis of answers to the reflection questions from the control group can give us some hints on this. Participants mentioned various outcomes from pressure. Some were related to the product or the process, such as an increase in product failure/bugs leading to a lack of trust in the product, product instability, neglect of long-run maintenance and testing activities, neglect of good practices, and overall lower quality. Some other outcomes were related to the client, such as unmet expectations and needs. Other outcomes were related to the team and estimators: overtime work, emotional distress, resignation, and solution block. Our supplementary material \cite{matsubara_moving_2023} presents quotations supporting each of these outcomes. Improvements in any of these outcomes can benefit individual practitioners and their companies. 

\begin{table}[!h]
\begin{tabular}{|p{8.4cm}|}
\hline
\textbf{Takeaway message:} The digital simulation and the study of the defense lenses is a low-time intervention with the potential to impact varied outcomes from pressure related to product/process quality, the client's needs, and the software practitioners' quality of life. \\ \hline
\end{tabular}
\end{table}

\section{Threats to Validity}
\label{sec:threats}

A threat to conclusion validity regards the reliability of measures. We performed a reliability analysis for TPB items after the first moment of data collection (pre-questionnaire, 45 data points). We got acceptable reliability scores (Cronbach's $\alpha$ higher than 0.7) for subjective norms and intentions. Nevertheless, we got lower values for attitudes (Cronbach's $\alpha$ of 0.66) and perceived behavioral control (Cronbach's $\alpha$ of 0.33). For attitudes, dropping one item resolves the issue. Dropping one item also improves subjective norms' and perceived behavioral control' Cronbach's $\alpha$. But to improve it further, we would need to add more items to the questionnaire and another step of data collection with the new questionnaire to compare answers before and after the interventions. This would require more time from participants in our study, potentially increasing mortality because software practitioners have a short time to devote to participation in research. As we valued keeping a sample of participants who were active software practitioners, increasing the relevance of our results, we decided to keep the questionnaire untouched during data collection. We dropped one item for each intention's antecedents during the final analysis, increasing their Cronbach's $\alpha$.

A threat to internal validity in our study regards mortality, as we had participants dropping out after the first moment of data collection: six in the experimental group and seven in the control group. Therefore, we compared dropouts to participants regarding demographic and TPB variables. We found significant differences in years of experience in software development and maintenance: dropouts were more experienced on average. We do not consider this an issue, as our proposal will likely benefit more inexperienced people.

Regarding construct validity, participants might not be familiarized with what is a defensive behavior toward software estimates. Moreover, we needed to design valid questions for assessing the TPB variables. Therefore, we constructed the questions about TPB with the guidance of a manual \cite{francis_constructing_2004}, which also required us to define to participants what is a defensive behavior of software estimates. We also piloted the questionnaires with four active software practitioners and improved the wording of questions to increase validity.

Controlled experiments, like ours, are generally limited in the number of subjects \cite{stol_abc_2018} and do not necessarily support statistical generalization \cite{baltes_sampling_2022}. Also, as typical in experiments, we prioritized internal over external validity. We valued assessing whether negotiation principles present in the defense lenses and the digital simulation could cause software practitioners to raise intentions in defending their software estimates over generalizing our results to a larger population. Nevertheless, the positive qualitative feedback we received from participants in the digital simulation can be seen as a preliminary sign of its external applicability in the software industry. 

\section{Conclusions}
\label{sec:conclusions}

In this paper, we contribute to the Software Engineering literature by evaluating a new strategy for practitioners to move on from the software engineers' gambit, in which they sacrifice quality---of products and of life---to gain time due to pressure over their software estimates. The proposed strategy is comprised of a set of lenses to support software estimators in defending software estimates during their communication to other relevant stakeholders. The lenses embed the principles of consolidated negotiation methods, thus also contributing to a concrete approach toward increasing negotiation skills among software practitioners. Moreover, we presented the defense lenses in a lightweight digital simulation format, making the acquisition of their knowledge more dynamic, low-cost, and requiring low time---important features for dissemination among already too-pressed software practitioners, who are used to play the software engineers' gambit. 

In addition, we provide supporting evidence that engaging with the digital simulation and learning the defense lenses increases participants' intentions in defending software estimates when facing pressure and their attitudes and perceptions of how good such behavior is from the perspective of other relevant people in their context. It also increases their choices of defense actions in comparison with a control group, revealing they are more likely to stop yielding to pressure. Furthermore, qualitative evidence shows that participants exposed to our approach found the principles can be useful in their daily industrial practice, further revealing the relevance of this work. Currently, we provided the booklet and the digital simulation for all participants in the study, including the ones in the control group. As part of our future work, we plan to follow up with participants to understand whether they applied the lenses in their work environments and their perception of their usefulness in the wild. 

\section{Data Availability}

The supplementary material with data that support the findings of this study is openly available in Figshare at \url{https://doi.org/10.6084/m9.figshare.20736844.v1}.

\section*{Acknoledgment}
    The present work is the result of the Research and Development (R\&D) project 001/2020, signed with Federal University of Amazonas and FAEPI, Brazil, which has funding from Samsung, using resources from the Informatics Law for the Western Amazon (Federal Law nº 8.387/1991), and its disclosure is in accordance with article 39 of Decree No. 10.521/2020. It also is supported by Universidade Federal do Amazonas (UFAM), Universidade Federal do Mato Grosso do Sul (UFMS), CAPES - Financing Code 001, CNPq processes 314174/2020-6 and 313067/2020-1, FAPEAM process 062.00150/2020, and grant \#2020/05191-2 S\~{a}o Paulo Research Foundation (FAPESP).

\bibliography{icse23}
\bibliographystyle{IEEEtranN}

\end{document}